\newcommand{\diracslash}[1]{#1\llap{/\kern2pt}}

\newcommand{\be}{\begin{equation}}
\newcommand{\ee}{\end{equation}}
\newcommand{\bea}{\begin{eqnarray}}\index{\footnote{}}
\newcommand{\eea}{\end{eqnarray}}
\newcommand{\ba}[1]{\begin{array}{#1}}
\newcommand{\ea}{\end{array}}

\documentclass[12pt]{iopart}
\usepackage{graphicx}
\usepackage[top=.5in, bottom=.5in, left=.8in, right=.8in]{geometry}
\begin{document}

\title{Number-phase uncertainty and quantum dynamics of  bosons and fermions  interacting with a finite range and large scattering length in a double-well potential }
\author{Kingshuk Adhikary$^1$, Subhanka Mal$^1$, Bimalendu Deb$^{1,2}$, Biswajit Das$^1$, Krishna Rai Dastidar$^1$ and Subhasish Dutta Gupta$^3$ }
  \address{$^1$ Department of Materials Science, Indian Association for the Cultivation of Science, Jadavpur, Kolkata 700032, India. $^2$ Raman Centre for Atomic, Molecular and Optical Sciences,
Indian Association for the Cultivation of Science, Jadavpur, Kolkata 700032, India. $^3$ School of Physics, University of Hyderabad, Hyderabad 500 046, India.}

\begin{abstract}

In a previous paper [Das B {\it et al.} J. Phys. B: At. Mol. Opt. Phys 2013 {\bf 46} 035501], it was shown that the unitary quantum phase operators play a 
particularly important role in quantum dynamics of bosons and fermions in a one-dimensional double-well (DW)  when the number of particles is small. In this paper, we define the standard quantum limit (SQL) for phase and number fluctuations, and describe two-mode squeezing  for number and phase variables. The usual two-mode number squeezing parameter, also used to describe two-mode entanglement of a quantum field, is defined considering phase as a classical variable. However, when phase is treated as a unitary quantum-mechanical operator, number and phase operators satisfy an uncertainty relation. As a result, the usual definition of number squeezing parameter becomes modified. Two-mode number squeezing occurs when the number fluctuation goes below the SQL at the cost of enhanced phase fluctuation. As an application of number-phase uncertainty,  we consider bosons or fermions trapped in a quasi-one dimensional double-well (DW) potential interacting via 
a 3D finite-range two-body  interaction potential with large scattering length $a_s$. Under tight-binding or two-mode approximation, we describe in detail the effects of the range of interaction 
on the quantum dynamics and number-phase uncertainty in the strongly interacting or unitarity regime $a_s \rightarrow \pm \infty$. Our results show intriguing coherent dynamics of number-phase uncertainty with number-squeezing for bosons and phase squeezing for fermions. Our results may  be important for exploring new quantum interferometry, Josephson oscillations, Bose-Hubbard and Fermi-Hubbard physics with ultracold atoms in DW potentials or DW optical lattices. Particularly interesting will be the question of the importance of quantum phase operators in two-atom interferometry and entanglement. 

\end{abstract}

\newpage

\section{Introduction}

With the advancement of research in the matter-waves of ultracold atoms in recent times, a new field called ``atom-optics'' has emerged \cite{book:meystre,shin:prl:2004,rmp:atom-interferometry}. A prototypical system to demonstrate interference between the atomic matter-waves is a double-well (DW) trap  where the macroscopic wave functions of two matter waves trapped in the lowest states of the two wells posses a well-defined  phase-difference which plays an essential role in many coherent effects such  the matter-wave interference \cite{ketterle:scienec:1997, pra:1997:walls} and Josephson oscillations in both atomic Bose \cite{prl:1997:giova,josephson:bosons1,josephson:bosons2} Fermi superfluids \cite{josephson:fermions}. Higher order coherence that underlies photon-photon interference or correlation as in well-known Hanbury Brown-Twiss effect has also been demonstrated with cold atoms \cite{hbt:aspect}. A matter-wave analogue of Hong-Wu-Mandel effect has also been realised with matter-waves \cite{aspect:2015}. Though there is a lot of parallelism between optics and ultracold matter-waves,
 there are some fundamental differences. First, the matter-waves have intrinsic nonlinearity that results from interactions between or among the atoms. Second, for a closed  matter-wave system such as DW trap, the total number of quanta (atoms) is conserved unlike that in optics. In quantum optics, unitary phase operators were introduced in 1980s by Barnett and Pegg \cite{pegg-burnett} to describe the phase measurement and quantum phase-dependent effects. The unitarity of  phase operators had remained an unresolved issue \cite{unitarity:dirac,unitarity1,unitarity2,unitarity3} for long since Dirac's seminal work in 1927 \cite{unitarity:dirac}. A series of pioneering experiments to measure quantum phase of optical fields was carried out by Mandel and coworkers in 1990s \cite{mandel:1991,mandel:1992,mandel:1993,mandel:1994}. A matter-wave counterpart of unitary phase operators has been introduced four years ago \cite{jpb:2013:biswajit}. Though unitary phase operators have attracted a lot of interests, they are 
yet to find wide-ranging applications. In defining unitary phase-difference operators for a two-mode field in optics, it is assumed that the total number of photons in the field is conserved. But in optics this assumption can 
not be fulfilled in general, except in closed quantum optical systems such as two-mode Raman type processes in high-Q cavities \cite{bdeb:1993:pra}.  But for matter-waves of ultracold atoms in a double-well trap, the total number of atoms is conserved during the trap lifetime or duration of any  experimental measurement on the trapped atoms. So, it is important to study the influence of unitary phase operators in matter-wave quantum dynamics. This will provide a new perspective in quantum optics with matter waves.

Motivated by the recent experimental and theoretical developments with  ultracold atoms in a tailor-made DW potential  \cite{selim:prl:2015,prl:2016:ketterle,transport3,transport2,transport1,transport4,transport5,transport10,transport11}; and by the prospect of  intriguing and controllable quantum effects such as quantum entanglement \cite{dwell:entanglement1,dwell:entanglement2,dwell:entanglement3} and  quantum  transport 
\cite{transport6,transport7,transport8,transport9,prl:2015:bodhaditya} using DW traps and DW optical lattices \cite{DWlattice1,DWlattice2,DWlattice3}, we here carry out a detailed model 
study on the quantum dynamics and quantum phases of few interacting bosons or fermions in a  DW potential, in terms of newly introduced quantum phase operators \cite{jpb:2013:biswajit} for matter-waves.  
The purpose of this work is to show  the effects of the finite-range of two-body interaction with large scattering length  on the quantum dynamics and quantum phase properties of bosons or fermions trapped in a double-well (DW) potential. This study may be important for 
Josephson effect of matter-waves in a DW potential. In Josephson oscillations,  the phase-difference between macroscopic        
wave functions of the two ensembles of particles residing on both sides of the Josephson junction is dynamically coupled with the difference in populations of the two ensembles. To describe Josephson phenomena, both the phase- and population-difference are usually treated as a pair of canonically conjugate classical dynamical variables. It remains an open question how unitary quantum phase operators will affect the Josephson effect \cite{book:pitaevskii:stringari}. Though here we do not address this question, our present study may serve as a precursor towards that direction. 

Here we study the effects of large scattering length and the finite range of interaction on
quantum dynamics and quantum phase fluctuations of few bosons or two or three fermions trapped in the double-well potential. We
consider Barnett-Pegg  type quantum phase operators for matter-wave of bosons or fermions \cite{jpb:2013:biswajit}. The
quantum phase operators for electromagnetic fields or photons is well-known. Quantum phase operators for matter-waves
are yet to attract research interest of the community working in the emerging areas of atom-optics. In order to regard a DW potential as
a basic building block for quantum atom-optical studies, it is important to understand from a fundamental point of view how quantum
nature of phases can affect the quantum phase fluctuations and number-phase uncertainty which are critical for Josephson effects. In this context, a DW potential can act as
a paradigmatic model. Our results show that, for low number of bosons or fermions quantum nature of the phase-difference between
the two localized states (sites) is quite important. This emerges from the dominant effect of the quantum vacuum state of the system on the
phase and number fluctuations. However, in the limit of large number of particles, the effect of the vacuum state becomes insignificant
and the results calculated using quantum phase operators reduce to those for classical phases as noted earlier \cite{jpb:2013:biswajit}.

We consider a three-dimensional trapping potential with a model symmetric  double-well structure along the axial direction ($z$-axis) and tightly confined harmonic trapping potentials along the transverse ($x$- and $y$-) directions. We assume that the temperature is low enough so that the atoms remain occupying the ground state of the radial harmonic potentials even in the strong atom-atom interaction regime at which the $s$-wave scattering length diverges. By integrating over the radial ground state, we reduce the  Hamiltonian into an effective one-dimensional (1D) model for double-well potential. In order incorporate an effective 1D atom-atom interaction, we consider two-parameter model atom-atom interaction potentials \cite{ijmp:2015:Bdeb,partha:injp:2015,partha:physscr:2016} with the parameters being the range $r_0$ of the interaction and $a_s$. These potentials  are capable of describing correctly  the unitarity regime ($a_s \rightarrow \pm \infty$) without any need for regularization. This is important 
because it is noted \cite{partha:injp:2015} that the analytical results of regularized contact interaction potential for two trapped atoms given by  Englert {\it et al.} in a seminal paper \cite{englert:foundphys}  can be reproduced  by taking the limit $r_0 \rightarrow 0$ in the case of an isotropic trap,  but not in case of quasi-1D in general.  
We calculate single-particle eigenvalues and eigenfunctions of this potential using discrete variable representation (DVR)-based \cite{acp:2003:Light}
Fourier Grid Hamiltonian (FGH) method \cite{jcp:1989:Ballint}. Using  these eigenvalues
and eigenfunctions, we  build a tight-binding model for interacting bosons or fermions
in the double-well. Tight-binding approximation makes use of localized (site-specific) basis constructed using the lowest two energy
eigenstates of the potentials. As a result, while any number of bosons can occupy the two lowest eigenstates, for two-component fermions
at most four fermions can be involved in the tight-binding model. Interestingly, we find that, even a small finite range induces an appreciable inter-site interaction apart from the on-site interaction.
In contrast, a contact potential gives rise only on-site interaction. 
The finite range of interaction between cold atoms becomes particularly
important for narrow Feshbach resonances. A magnetic Feshbach resonance (MFR) \cite{rmp:2010} or its optical counterpart optical Feshbach resonance
(OFR) \cite{fedichev:prl:1996,bdeb:prl:2009} has been an important tool for altering $a_s$ over a wide range from large negative to large positive values, and thereby to achieve
strongly interacting ultra-cold atomic gases.

Here we summarise the main findings of this study. First, the effective on-site interaction varies nonlinearly with $r_0$ with a maximum the position of which depends on the length scale of the transverse harmonic oscillator. The tighter the transverse confinement the lower is the value 
of $r_0$ at which maximum occurs. For relatively large $r_0$, one can not ignore the inter-site interaction. Since the pair-tunneling probability 
increases with the increase of $U_0$, the pair-tunneling probability exhibits non-monotonic behavior with the variation of $r_0$. Since the on-site interaction affects the tunneling probability, it is expected that 
the expectation of phase and number operators and their fluctuations will depend non-monotonically on $r_0$. 
Second,  in the case of two atoms trapped in the double well, either  the single-particle or  two-particle (pair) tunneling probability  will dominate 
over the other depending on the initial condition. In the Josephson picture, quantum tunneling oscillations  or temporal variation of the number-difference is driven by the current arising from the spatial gradient of the phase-difference. For macroscopically large number of particles, the Josephson oscillations can be interpreted with a classical description where the phase and the number are treated as classical variables which are canonically conjugate. Now, the question arises: Is there any two-particle or few particle analogue of Josephson oscillations? We here assert that in such few particle Josephson-like  oscillations, it is essential that the phase and number variables are described in terms of operators. Unitary phase operators for few-particle systems, the quantum vacuum state plays an essential and important role. Our results show that the single- and two-particle tunneling probabilities are intertwined with the quantum fluctuations of number and phase operators. Third, the results on average 
values of number and phase operators  for two 
bosons and a pair of two-component fermions are found to be similar, however their fluctuation properties differ. Fourth, the temporal evolution of the expectation values of  number and phase operators and their fluctuations show intriguing collapse and revival dynamics the origin of 
which  can be traced to the nonlinearity due to atom-atom interactions. With increasing number of bosons, the collapse and revivals exhibit multiple time scales and modulations due to increase in number-density-dependent nonlinear term. Fifth, there is number squeezing in bosons depending on the system parameters and initial conditions, but no phase squeezing in bosonic systems. In contrast, there are both number and significant phase squeezing for a pair of two-component fermions. The phase squeezing in case of two fermions may be related to fermionic exchange symmetry and associated inherent two-particle entanglement property.

The paper is organized in the following way. To  start with, we first recapitulate the two-mode unitary quantum phase-difference  operators that 
are canonically conjugate to the number-difference operator of the two modes in the next section. We also discuss the commutation relations among these operators. In section 3, we consider an one-dimensional (1D) DW  potential derived from a 3D trap potential. We calculate single-particle 
eigenvalues and eigenfunctions for this potential using DVR-based FGH method. We then develop a tight-binding model for a fixed number of interacting bosons or fermions. We consider that the two-body interaction is finite-ranged and can take into account $s$-wave Feshbach resonances. Our numerical results are discussed and analysed in detail in section 4. The paper is concluded in section 5.

\section{Quantum phase operators} 

\subsection{Bosonic phase-difference operators}\label{2.1}

In this section,  we first give a brief outline on two-mode phase difference and number-difference operators for Bosonic particles. Two operators $\hat{C}_{12}$ 
and $\hat{S}_{12}$ corresponding to the cosine and sine, respectively,  of the phase-difference are defined, where the subscripts `1' and `2' refer to the two modes of the bosonic field.  Neither $\hat{C}_{12}$ nor $\hat{S}_{12}$ commutes with the number-difference operator $\hat{W}_{12}$.  As a result, there are two uncertainty relations 
corresponding to the products between the quantum fluctuations of $\hat{C}_{12}$ and $\hat{W}_{12}$ and between those of  $\hat{S}_{12}$ and $\hat{W}_{12}$. We next combine these two uncertainty relations into one and define SQL for fluctuation in phase-difference or number-difference operator. 

Carruthers and Nieto \cite{unitarity3} defined two Hermitian phase operators $\hat C$ for cosine of and $\hat S$ for sine of quantum phase of an optical field, 
but they are non-unitary. Using these operators, they introduced two-mode phase difference operators 
\begin{eqnarray}
\hat C^{CN}_{12} = \hat C_1\hat C_2 + \hat S_1\hat S_2 \nonumber \\
\hat S^{CN}_{12} = \hat S_1\hat C_2 - \hat S_2\hat C_1 
\end{eqnarray}
where
\begin{eqnarray}
\hat C_i=\frac{1}{2}[(\hat N_i+1)^{-\frac{1}{2}} \hat a_i+\hat a^\dagger_i(\hat N_i+1)^{-\frac{1}{2}}] 
\eea
\bea 
\hat S_i=\frac{1}{2i}[(\hat N_i+1)^{-\frac{1}{2}}\hat a_i-\hat a^\dagger_i(\hat N_i+1)^{-\frac{1}{2}}]
\end{eqnarray}
are the phase operators corresponding to the cosine and sine,  respectively, of $i$-th mode, where $\hat a^\dagger_i$($\hat a_i$) denotes the creation(annihilation) operator for a boson  
and $\hat{N}_i = \hat{a}_i^{\dagger} \hat{a}_i$. The explicit
form of phase-difference operators can be written (with $i$=1 or 2) as 
\begin{eqnarray}
\hat C^{CN}_{12} = \frac{1}{2}[(\hat N_1+1)^{-\frac{1}{2}}\hat a_1\hat a^\dagger_2(\hat N_2+1)^{-\frac{1}{2}}+  \hat a^\dagger_1(\hat N_1+1)^{-\frac{1}{2}}(\hat N_2+1)^{-\frac{1}{2}}\hat  a_2] 
\eea
\bea 
\hat S^{CN}_{12} = \frac{1}{2i}[(\hat N_1+1)^{-\frac{1}{2}}\hat a_1\hat a^\dagger_2(\hat N_2+1)^{-\frac{1}{2}}- \hat a^\dagger_1(\hat N_1+1)^{-\frac{1}{2}}(\hat N_2+1)^{-\frac{1}{2}}\hat a_2]
\end{eqnarray}
The above operators are non-unitary. Burnett and Pegg \cite{pegg-burnett} first introduced a Hermitian and unitary phase operator. As shown in Ref.\cite{bdeb:1993:pra}, following Barnett-Pegg formalism, one can define unitary operators corresponding to cosine
and sine of phase-difference by coupling vacuum state of one mode and highest Fock state to another mode in a finite dimensional Fock space resulting in the expressions 
\begin{eqnarray}
\hat C_{12} = \hat C^{CN}_{12} + \hat C^{(0)}_{12} \label{eq6} \\
\hat S_{12} = \hat S^{CN}_{12} + \hat S^{(0)}_{12}
\label{eq7}
\end{eqnarray}
where $N = \langle \hat{N}_1 \rangle + \langle \hat{N}_2 \rangle$ is total number of bosons which is conserved and 
\begin{eqnarray}
\hat C^{(0)}_{12} = \frac{1}{2}[|N,0\rangle \langle 0,N|+|0,N\rangle \langle N,0|]\\
\hat S^{(0)}_{12} =  \frac{1}{2i}[|N,0\rangle \langle0,N|-|0,N\rangle \langle N,0|]
\end{eqnarray}
are the operators that are constructed by  coupling the vacuum state of one mode with the highest Fock state of the other mode. 
$|N_1,N-N_1\rangle$ represents a two mode Fock state with $N_1$ photons in mode 1 and remaining $(N-N_1)$ in mode 2.
The difference of the number or the population imbalance between the two wells is $\hat W=\hat N_1-\hat N_2$.
The commutation relations of the given operators $\hat C_{12},\hat S_{12}$ and $\hat W$  are as follows  
\begin{equation}
 [\hat {C_{12}}, \hat {W}]=2i(\hat S_{12}-(N+1)\hat S^{(0)}_{12})\nonumber
\end{equation}
\begin{equation}
 [\hat {S_{12}}, \hat {W}]=-2i(\hat C_{12}-(N+1)\hat C^{(0)}_{12})\nonumber
\end{equation}
\begin{equation}
 [\hat C_{12},\hat S_{12}]=0
\end{equation}
The first two of the above equations imply 
\begin{eqnarray}
{\Delta C_{12}} \Delta W \ge  \left |  S_{12}  - (N+1)  S^{(0)}_{12} \right | \label{eq13} \\
{\Delta S_{12}} \Delta W \ge  \left |  C_{12}  - (N+1)  C^{(0)}_{12} \right | 
\label{eq14}
\end{eqnarray}
where $\Delta A =\sqrt{\langle \hat{A}^2 \rangle - \langle\hat A \rangle^2}$ is the fluctuation for the operator $\hat{A}$ and $A = \langle \hat{A} \rangle$ is the expectation 
of $\hat{A}$. We define the normalized number-difference operator by 
\bea 
\hat{W}_n = \frac{\hat{N}_1 - \hat{N}_2 }{\langle \hat{N}_1 + \hat{N}_2 \rangle } = \frac{\hat{W}}{N}
\eea 
Now, squaring and summing the two inequalities (\ref{eq13},\ref{eq14}) and dividing the resultant inequality by $N^2$, we obtain
\begin{equation}
 ({\Delta C_{12}}^2+{\Delta S_{12}}^2) {\Delta W_n}^2 \ge \frac{1}{N^2} [[S_{12}  - (N+1)  S^{(0)}_{12}]^2+[ C_{12}  - (N+1)  C^{(0)}_{12}]^2]
\end{equation}

Now, we define the standard quantum limit of fluctuation $\Delta_{SQL}$ in number-difference or phase-difference quantity by 
\bea 
\Delta_{{\rm SQL}} = \frac{1}{N}\sqrt{[S_{12}  - (N+1)  S^{(0)}_{12}]^2+[ C_{12}  - (N+1)  C^{(0)}_{12}]^2}
\eea 
We define the normalized squeezing parameters for both phase- and number-difference operators,  respectively,  by  
\begin{equation}
\Sigma_p={\Delta E_{\phi}}^2- \Delta_{{\rm SQL}} 
\end{equation}
and 
\begin{equation}
\Sigma_w={\Delta W_n}^2- \Delta_{{\rm SQL}} 
\end{equation}
where $\Delta E_{\phi} = \sqrt{({\Delta C_{12}})^2 + ({\Delta S_{12}})^2 } $ is an average phase fluctuation. The system will be squeezed in number or phase variables whenever 
$\Sigma_w$ or $\Sigma_p$, respectively, becomes negative.

When the phase is treated as a classical variable, the entanglement in number variables between
the two modes is described by the two-mode squeezing \cite{mtqo:1997:Radmore-Barnett}
or entanglement \cite{pra:2002:Deb} parameter given by 
\bea
\xi_n = (\Delta(\hat N_1 - \hat N_2))^2/N
= \frac{\left ( \Delta\hat{W} \right )^2 }{N}
\eea
The two modes are said to be entangled in number variables when $\xi_n < 1$ or if  $\Delta\hat{W}$ is less than $\sqrt{N}$. Instead of $\xi_n$, one can quantify the entanglement 
by the parameter $\Sigma_{n} = (\xi_n -1)/N $ the negativity of which will imply entanglement between the modes. 
For large number of bosons, 
$ C^{(0)}_{12} \rightarrow 0$, $S^{(0)}_{12} \rightarrow 0$  and $(C_{12} + S_{12}) \rightarrow 1$, that is,  the vacuum fluctuations become negligible. In these limits, it is easy to see that $\Sigma_w \rightarrow \Sigma_n$.

\subsection{Fermionic phase difference operators}
We consider spin-half fermions in a DW potential. 
In the tight-binding approximation, a single mode of fermionic matter-wave is characterized by the spin component $(\sigma=\uparrow$ or $\downarrow)$ and the site index $s=l$ (left) or r(right) or equivalently index 1 or 2. Owing to Pauli's exclusion principle, at most four two-component fermions can be placed in the two lowest energy eigenstates of DW potential, or equivalently to the two sites of the potential under TBA. To study coherent fermionic quantum dynamics under TBA, we consider two cases: In case-1 we study a pair of interacting two-component fermions, and in case-2 we discuss the dynamics of three such fermions. The state with four fermions is a trivial one since all the available states under TBA are filled up and so there will be no dynamical evolution.  

To define a phase operator for spin-half fermion systems, let us first consider a general multi-particle fermion system, not necessarily in a DW potential but in any general geometry.  
Because of Pauli's exclusion principle, a single mode can be occupied at most by a single fermion. So, a many-fermion system will obviously be a multi-mode system. Then we need to consider a fermion  phase-difference  operator between two spatial modes out of many modes. Let us characterize  any two chosen 
modes by the symbols  ($ l$) and ($r$) where `$l$' and `$r$' denote any spatial modes (not necessarily `left' and `right'). Let  $\sigma$ and $\sigma'$ represent either spin up ($\uparrow$) or spin down ($\downarrow$).   
The general form of two-mode fermionic sine and cosine phase-difference operators \cite{jpb:2013:biswajit} are then given by
\begin{eqnarray}
\hat{C}^F_{l\; r} &=& \frac{1}{{\mathcal N }_c} \sum_{\sigma \sigma' }  \left [ \frac{1}{2}\left \{ (\hat{N}_{l\sigma}+1)^{-\frac{1}{2}} \hat{a}_{l\sigma} \hat{a}^{\dagger}_{r\sigma'}(\hat{N}_{r\sigma'}+1)^{-\frac{1}{2}} + (\hat{N}_{r\sigma'}+1)^{-\frac{1}{2}} \hat{a}_{r\sigma'} \hat{a}^{\dagger}_{l\sigma}(\hat{N}_{l\sigma}+1)^{-\frac{1}{2}} \right \} \right ]  \nonumber\\   &+& 
 \frac{1}{{\mathcal N}_c} \sum_{\sigma \sigma' } \left [ \frac{1}{2}\sum_{jk} \left \{ |10\rangle _{j\hspace{.05in} k}\langle 01| + |01\rangle _{k\hspace{.05in} j}\langle 10|\right \} \right ],  
\eea
\bea
\hat{S}^F_{l\; r}&=& \frac{1}{{\mathcal N}_c } \sum_{\sigma \sigma' }  \left [  \frac{1}{2i} \left \{ (\hat{N}_{l\sigma}+1)^{-\frac{1}{2}} \hat{a}_{l\sigma} \hat{a}^{\dagger}_{r\sigma'}(\hat{N}_{r\sigma'}+1)^{-\frac{1}{2}} - (\hat{N}_{r\sigma'}+1)^{-\frac{1}{2}} \hat{a}_{r\sigma'} \hat{a}^{\dagger}_{l\sigma}(\hat{N}_{l\sigma}+1)^{-\frac{1}{2}} \right \} \right ] \nonumber\\ &+& 
\frac{1}{{\mathcal N_c}} \sum_{\sigma \sigma' }  \left [ \frac{1}{2i}\sum_{jk} \left \{ |10\rangle _{j\hspace{.05in} k}\langle 01| - |01\rangle _{k\hspace{.05in} j}\langle 10| \right \} \right ]  
\end{eqnarray}
Where $|01\rangle_j$ represents $j$-th combination of states where $l\sigma$ state is occupied  but $r\sigma'$ is empty and $|10\rangle_k$ represents $k$-th combination of states with $l\sigma$ state being occupied  and $r\sigma'$ empty. Here ${\mathcal N}_c$ is the total number of spin configurations in the two spatial modes. For instance, for a pair of spin-half fermions - one in $\uparrow$ state and the other in $\downarrow$ state, ${\mathcal N}_c =2$, while for 3 spin-half fermions with any two either in $\uparrow$ or $\downarrow$ and the other in $\downarrow$ or $\uparrow$, respectively, we have ${\mathcal N}_c = 4$. 

The fluctuation of fermionic operators are defined as
\begin{eqnarray}
\Delta {C}_{l \hspace{0.02in} r} = \sqrt{\langle \hat{C}^2_{l \; r}\rangle - \langle \hat{C}_{l \; r} \rangle ^2} 
\eea 
\bea
\Delta {S}_{l \; r} = \sqrt{\langle \hat{S}^2_{l \; r}\rangle - \langle \hat{S}_{l \; r} \rangle ^2} 
\end{eqnarray}  
The number difference operator is given by
\begin{eqnarray}
\hat{W}^F = \sum_\sigma(\hat{a}^{\dagger}_{l\sigma} \hat{a}_{l\sigma} - \hat{a}^{\dagger}_{r\sigma} \hat{a}_{r\sigma}) 
\end{eqnarray}
and the fluctuation is given by,
\begin{eqnarray}
\Delta{W}^F = \sqrt{\langle \hat{W^F}^{ 2}\rangle - \langle \hat{W}^F\rangle^2} 
\end{eqnarray}
The commutation algebra of the given operators have following form,
\begin{equation}
 [\hat{C}^F_{l \; r},\hat W^F] = -4i\hat{S}^F_{l \; r}
\end{equation}
\begin{equation}
 [\hat{S}^F_{l \; r},\hat W^F] = 4i\hat{C}^F_{l \; r}
\end{equation}
\begin{equation}
 [\hat{C}^F_{l \; r},\hat{S}^F_{l \; r}] = 0
\end{equation}
The first two of these commutators provide the number-phase uncertainty relations 
\bea 
\Delta C_{l \; r}^F \Delta W^F \ge 2|S_{l \; r}^F| \\
\Delta S_{l \; r}^F \Delta W^F \ge 2|C_{l \; r}^F|
\eea 
Now, following the same procedure as in bosons, and writing the normalized number-difference operator $\hat{W}^F_n = \hat{W}^F/N_F$, with $N_F = \sum_{\sigma, s=l,r} \langle \hat{a}_{\sigma,s}^{\dagger} \hat{a}_{\sigma,s} \rangle $ being the total number of fermions, we define the SQL of fluctuation for 
phase-difference and number-difference variables, respectively,  by  
\bea 
\Delta_{{\rm SQL}}^{F} = \frac{2}{N_F} \sqrt{ \left ( S_{l \;r}^F\right)^2 + 
\left( C_{l \; r}^F\right)^2}
\eea 
We the introduce the 
fermionic phase-difference and number-difference squeezing parameters by 
\bea 
\Sigma_p^{ F} = \left (\Delta E_{\phi}^{l \; r} \right )^2 - \Delta_{{\rm SQL}}^{ l \; r}
\eea 
\bea
\Sigma_w^{ F} = \left ( \Delta W^{F}_n \right )^2 - \Delta_{{\rm SQL}}^{ l \; r}
\eea 
where $\Delta E_{\phi}^{ l \; r} = \sqrt{\left(\Delta C_{l \; r}^F\right)^2 + \left(\Delta S_{l\; r}^F\right)^2 }$. Fermionic phase operators do not have any classical analogue or limit unlike that in bosons. Furthermore, fermionic number-phase uncertainty relations are significantly different from those of bosons.  

We next apply these bosonic and fermionic number-phase operator formalism to finite number of interacting bosonic and fermionic atoms in a DW potential to numerically illustrate the number-phase uncertainty and squeezing, their dynamical evolution, dependence on the range of interaction. In the next section, we describe in brief a model DW potential.

\section{Tight-binding DW models with finite-range interactions and their solutions}


\subsection{Model DW potential}
Let us consider a 3D trap potential $V_{trap}(r)$ with the following features: (i) along x- and y- axes, it is harmonic trap; (ii) along z axis it is double well-trap. We may write 
\begin{eqnarray}
V_{trap}(r)=V(\rho)+V(z)=\frac{1}{2}m{\omega_\rho}^2{\rho}^2+\frac{1}{2}\lambda^2(z^2-\eta^2)^2
\label{eq18}
\end{eqnarray}
where $\omega_\rho$ is radial frequency,  ${\pm\eta}$ are the two minimum points where the potential vanishes and the barrier height is $V_0=\frac{1}{2}\lambda^2\eta^4$. So, the parameter $\lambda^{2}$ has the dimension of energy-length$^{-4}$.
The barrier height and offset energy between two wells can be
dynamically controlled by the laser intensity and the relative phase between the lasers. If the barrier height between the wells is very
large compared to the ground-state energy, then for atoms occupying the lowest energy band of the double-well, each well becomes almost independent harmonic oscillator. Under such conditions, we calculate the harmonic oscillator ground-state energy and the width of the ground-state. We use the width as the length scale of the problem under tight-binding approximation.
However, we calculate exact single-particle energy eigenvalues and eigen-functions of a particle in a DW by FGH method. Under harmonic approximation the harmonic frequency $\omega_z=\frac{2\lambda\eta}{\sqrt{m}}$ and the harmonic oscillator length scale $a_l=\sqrt{\frac{\hbar}{m\omega_z}}$.

Here we convert 3D radial position $r$ into cylindrical coordinates $(\rho,z)$. Since $V_{trap}(r)$ is only radial we ignore the azimuthal angel $\phi$.
We may write $r=\sqrt{\rho^2+z^2}$. The dimensionless forms of the radial harmonic potential and the axial double-well potential are given by $\bar V(\bar\rho)=\frac{1}{2}\frac{\bar\rho^2}{a_{\rho}^2}$
and $\bar V(\bar z)=\frac{1}{2}\bar\lambda^2(\bar z^2-\bar\eta^2)^2$ respectively; where, $a_{\rho}=\frac{\omega_z}{\omega_{\rho}}$, $\bar z=z/a_{l}$ and $\bar\rho=\frac{\rho}{a_l}$ are dimensionless parameters.

\subsection{Finite-range interaction}
Now we describe the interaction between the two particles or atoms trapped in a DW.
At first we shall concentrate on finite range interaction and see the effects of range and scattering length. For this study we choose our model finite-range potential \cite{ijmp:2015:Bdeb} with large scattering length
\begin{eqnarray}
V_{int}(|{\bf r_1-r_2}|) = -\frac{\hbar^2 \kappa^2}{\alpha \mu} \frac{1}{\cosh^2(\kappa r)}
\label{eq22}
\end{eqnarray}
where $r = |{\bf r_1-r_2}|$, ${\bf r_1}$ and ${\bf r_2}$ represent the positions of two interacting particles 1 and 2, respectively, $\mu$ is the reduced mass, $r_0$ is the range, $a_s$ is scattering length, $\alpha = \sqrt{1-\frac{2r_0}{a_s}}$, $\beta = 1+\alpha$ and $\kappa = \frac{\beta}{r_0}$. For negative scattering length $\alpha$ is bounded
by 1$\leq $ $\alpha$ $<$ 2 and for positive scattering length $\alpha$ is bounded by 0 $<$ $\alpha$ $\leq $2. We take dimensionless wavefunction 
\begin{eqnarray}
\Phi(\bar r)=\frac{1}{\sqrt{a_{\rho}\pi}} e^{-\frac{\bar\rho^2}{2a_{\rho}}}\psi_{1D}(\bar z) 
\end{eqnarray} 
where $\psi_{1D}(z)$ is calculated numerically.
So, in terms of this localized basis, there will be three coefficients of interaction
\begin{eqnarray}
U_{i \;j}  = \int\int |\Phi_i ({\bf r_1})|^2 V_{int}(r) |\Phi_j ({\bf r_2})|^2 d{\bf r_1}d{\bf r_2}
\end{eqnarray}
where `$i$' and `$j$' stand for the site index `l' (left) and `r' (right), or equivalently `1' and `2'. 
As our potential (double well) is one dimensional (along z-direction), we reduce these coefficients in terms of one-dimensional wave functions i.e, the localized eigen states of the double-well potential 
\begin{eqnarray}
U_{i \;j} = -\frac{\hbar^2\kappa^2}{\alpha\mu a_l^2 a_\rho}\int \frac{e^{-\frac{\rho^2}{2a_l^2a_\rho}}|\psi_i(z_1)|^2 |\psi_j(z_2)|^2}{\cosh^2(\kappa\sqrt{\rho^2 +(z_1-z_2)^2})}\rho d\rho dz_1 dz_2 
\end{eqnarray}
Here we choose the frequency along radial direction$(\omega_\rho)$ to be at least one order larger than the frequency along $z$-axis$(\omega_z)$. For $|a_s|>> 2r_0$, the results have universal behavior in the sense that they do not depend on the sign and magnitude of $a_s$.
\begin{figure}[h!]
 \centering
 \vspace{.35in}
  \hspace{-.4in}
  \includegraphics[height=3.5in, width=6in]{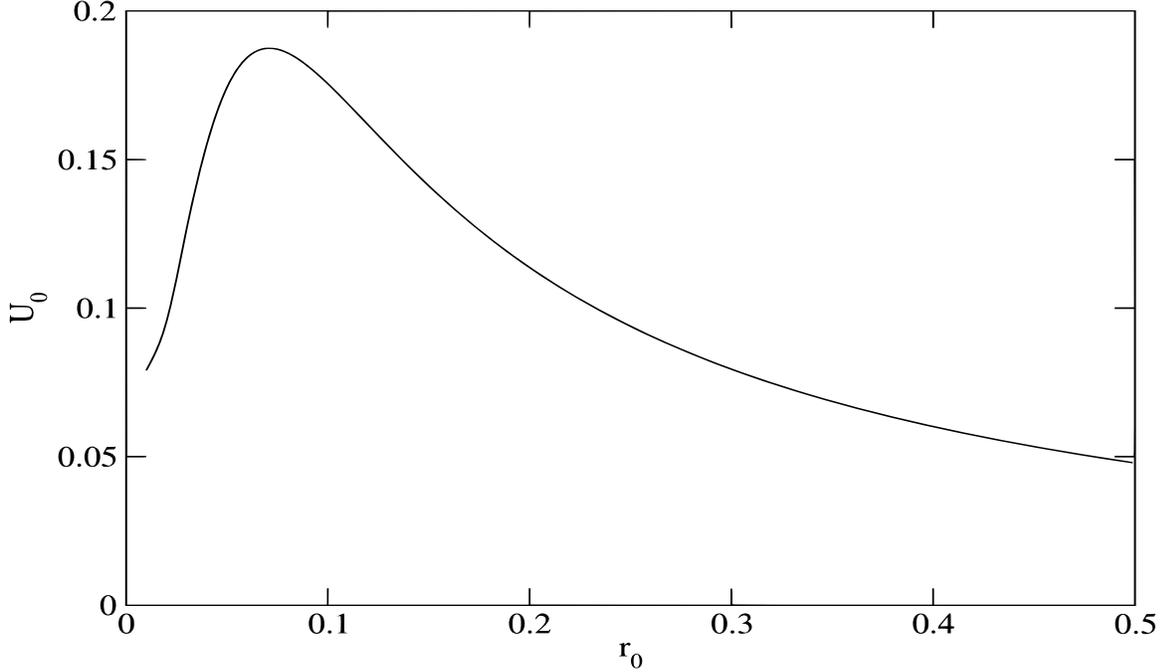}
 \caption{\small Variation of on-site interaction $U_0$ (in unit of $\hbar\omega_z$) with range $r_0$ (in unit of $a_l$) for large scattering length ($a_s = 100a_l$) with trap parameters $\omega_\rho = 10\omega_z$, $\bar\lambda=0.5$ and $\bar\eta=2$.}
 \label{Figure 1.}
  \end{figure}

\subsection{Bosons}

The Hamiltonian of a system of $N$ bosons in terms of the two lowest energy eigen basis of the DW is given by
\begin{eqnarray}
 \hat H_{eb} = \sum_{i=1}^{2}(\epsilon_{i} \hat{a}_{i}^{\dagger} \hat{a}_{i} + \frac{\hbar}{2}U_{i}\hat{a}_{i}^{\dagger}\hat{a}_{i}^{\dagger}\hat{a}_{i}\hat{a}_{i}) + \frac{\hbar}{2}\sum_{i\neq j}U_{ij}\hat{a}_{i}^{\dagger}\hat{a}_{j}^{\dagger}\hat{a}_{j}\hat{a}_{i} \nonumber
\end{eqnarray}
where, $\hat{a}_{i} (\hat{a}_{i}^{\dagger})$ is the bosonic annihilation(creation) operator, $\epsilon_{i}$ is the energy of the $i$-th energy state, $U_{i}$ is the interaction between two particles in the $i$-th state and $U_{ij}$ is the interaction between two particles in $i$-th and $j$-th states, respectively.
This Hamiltonian is written in energy-eigen basis, but one can write it, under tight-binding approximation, in terms of localized basis. 
The bosonic Hamiltonian for a finite-range  interaction in the localized  basis can be  written as
\begin{eqnarray}
\hspace{-.5in}
\hat {H}=-\hbar J(\hat a^\dagger_l\hat a_r+\hat a^\dagger_r\hat a_l)+\frac{\hbar}{2}U_l\hat a^{\dagger 2}_l\hat a^{2}_l+\frac{\hbar}{2}U_r\hat a^{\dagger 2}_r\hat a^{2}_r
  +\frac{\hbar}{2}U_{lr}\hat a^\dagger_l\hat a^\dagger_r\hat a_r\hat a_l +\frac{\hbar}{2}U_{rl}\hat a^\dagger_r\hat a^\dagger_l\hat a_l\hat a_r
\end{eqnarray}
where, $\hat{a}_{l} (\hat{a}_{l}^{\dagger})$ represents of bosonic annihilation(creation) operator in the left localized basis and $\hat{a}_{r} (\hat{a}_{r}^{\dagger})$ represents the same in right 
localized basis. Here $J>0$ is the tunneling term between two localized site, $U_l = U_{ll} (U_r = U_{rr})$ is on-site interaction in left(right) well and  $U_{lr}=U_{rl}$ is inter-site interaction. In general, $U_l$ and $U_r$ are different but in symmetric double well they are equal. 

Now, we develop wavefunction $|\psi(t)\rangle$ in Fock state basis with total number of Bosons (N) fixed.
The wavefunction $|\psi(t)\rangle$ is given by 
\begin{eqnarray}
|\psi(t)\rangle=\sum_{l=0}^{N}c_l(t)|l,N-l\rangle 
\end{eqnarray}
where, $c_l(t)$ is the probability amplitude for $l$ particles in left well and $(N-l)$ particles in right well. The Fock state basis $|l,N-l\rangle$ implies $l$ number particles occupied left well and $(N-l)$ number particles occupied right well. From time-dependent Schr\"{o}dinger equation
\begin{equation}
 i\hbar\frac{d}{dt}|\psi(t)\rangle=\hat{H}|\psi(t)\rangle \nonumber 
 \end{equation}
we get the recurrence relation which is,
\begin{eqnarray}
i\frac{dc_l}{dt}=-[c_{l-1}\kappa_{l-1}+c_{l+1}\kappa_{l}]+c_l[V_l+V_r+2V_{lr}]
\end{eqnarray}
where,
$\kappa_l=J\sqrt{(l+1)(N-l)}$,\hspace{0.1in}$V_l=\frac{U_l}{2}l(l-1)$,\hspace{0.1in}$V_r=\frac{U_r}{2}(N-l)(N-l-1)$,\hspace{0.1in}$V_{lr}=\frac{U_{lr}}{2}l(N-l)$.
Here the normalization condition is $\sum_{l=0}^N|c_l(t)|^2=1$.
The pair tunneling probability is given by $P_2 = |c_2|^2$ or $P_2 = |c_0|^2$ provided the both atoms are initially in the right or left well, respectively; while the single particle tunneling probability is $P_1 = |c_1|^2$ with both particles initially being in the either well. The expectation value of unitary cosine and sine phase-difference operators are
\begin{eqnarray}
C_{12}=C_{CN}+\frac{1}{2}[c_0^*c_N+c_N^*c_0]\\
S_{12}=S_{CN}+\frac{1}{2i}[c_N^*c_0-c_0^*c_N]
\end{eqnarray}
where,
\begin{eqnarray}
C_{CN}=\frac{1}{2}[\sum_{l=1}^Nc_{l-1}^*c_l+\sum_{l=0}^{N-1}c_{l+1}^*c_l]\\
S_{CN}=\frac{1}{2i}[\sum_{l=1}^Nc_{l-1}^*c_l-\sum_{l=0}^{N-1}c_{l+1}^*c_l]
\end{eqnarray}
Fluctuation of unitary cosine and sine phase operators are
\begin{eqnarray}
\Delta C_{12}=\sqrt{[C^2_{12}-{C_{12}}^2]} \\               
\Delta S_{12}=\sqrt{[S^2_{12}-{S_{12}}^2]}
\end{eqnarray}
The Expectation and fluctuation value of $\hat W$ are
\begin{eqnarray}
 W = \sum_{l=0}^N (2l-N)|c_l(t)|^2\\ 
\Delta W =\sqrt{[\sum_{l=0}^N |c_l(t)|^2(2l-N)^2-(\sum_{l=0}^N (2l-N)|c_l(t)|^2)^2]}
\end{eqnarray}

\subsection{ Two-fermion system}\label{3.4}
Now we are going to  study the dynamics of a few fermions in a symmetric double well. As we know that the fermions obey Pauli's principle, it is essential to take care of large number of single particle states even in the very low energy limit. Recently, quantum states of a pair of two-component fermions in a controllable DW potential has been experimentally prepared as a building block for two-site Hubbard model \cite{selim:prl:2015}. 
Considering the fermions are of two-component type and the interaction is finite-ranged, the many-body Hamiltonian is given by
\begin{eqnarray}
\hspace{-1in}
 \hat{H} = -J\hbar(\hat{a}^{\dagger}_{l\uparrow} \hat{a}_{r\uparrow} + \hat{a}^{\dagger}_{l\downarrow} \hat{a}_{r\downarrow} + \hat{a}^{\dagger}_{r\uparrow} \hat{a}_{l\uparrow} + \hat{a}^{\dagger}_{r\downarrow} \hat{a}_{l\downarrow}) + U_l \hbar \hat{a}^{\dagger}_{l\uparrow} \hat{a}^{\dagger}_{l\downarrow} \hat{a}_{l\downarrow} \hat{a}_{l\uparrow} + U_r \hbar \hat{a}^{\dagger}_{r\uparrow} \hat{a}^{\dagger}_{r\downarrow} \hat{a}_{r\downarrow} \hat{a}_{r\uparrow} \nonumber\\+ U_{lr} \hbar(\hat{a}^{\dagger}_{l\uparrow} \hat{a}^{\dagger}_{r\uparrow} \hat{a}_{r\uparrow} \hat{a}_{l\uparrow}\nonumber
 + \hat{a}^{\dagger}_{l\downarrow} \hat{a}^{\dagger}_{r\uparrow} \hat{a}_{r\uparrow} \hat{a}_{l\downarrow} + \hat{a}^{\dagger}_{l\uparrow} \hat{a}^{\dagger}_{r\downarrow} \hat{a}_{r\downarrow} \hat{a}_{l\uparrow} + \hat{a}^{\dagger}_{l\downarrow} \hat{a}^{\dagger}_{r\downarrow} \hat{a}_{r\downarrow} \hat{a}_{l\downarrow}) 
\end{eqnarray}
The trial wave function for this case,
\begin{eqnarray}
|\psi(t)\rangle = \frac{c_1(t)}{\sqrt{2}}(|\uparrow , \downarrow\rangle + |\downarrow , \uparrow\rangle) +c_2(t)|\uparrow\downarrow , 0\rangle +c_3(t)|0 ,\uparrow\downarrow\rangle \nonumber
\end{eqnarray}
The state $|\uparrow , \downarrow\rangle$ represents one fermion (with up spin) is in the left well and one fermion (with up spin) is in the right well and $\frac{|C_1(t)|^2}{2}$ is the probability of finding the state and so on for the other states. Solving the Schr\"{o}dinger equation we get the linear coupled differential equations:
\begin{eqnarray}
i\frac{dc_1}{dt} &=& -\sqrt{2}J(c_2 + c_3) + U_{lr}c_1 \nonumber\\
i\frac{dc_2}{dt} &=& -\sqrt{2}Jc_1 + U_lc_2 \\
i\frac{dc_3}{dt} &=& -\sqrt{2}Jc_1 + U_rc_3 \nonumber
\end{eqnarray}
For the initial conditions $c_1(0)=1$, $c_2(0)=0$, $c_3(0)=0$, the analytical solutions are given by 
\begin{eqnarray}
c_1(t) &=&  \frac{e^{-i\frac{\bar U\tau}{2}}}{4 \bar{\Omega}}  \left [ \left ( \bar{U'}+ 2\bar{\Omega} \right ) e^{i \bar{\Omega}\tau} - \left (\bar{U'}- 2 \bar{\Omega} \right ) e^{-i\bar{\Omega}\tau}  \right ]\nonumber\\
c_2(t) &=& c_3(t)
= \sqrt{2}i \frac{e^{-i\frac{\bar U\tau}{2}}}{\bar{\Omega}}  \sin(\bar{\Omega}\tau)
\end{eqnarray}
where $\bar{\Omega} = \sqrt{(\bar{U'}/2)^2 + 4} $. 
 Here $U_l=U_r=U$, $\bar{U}=U/J$, $\bar{U'}=\frac{U-U_{lr}}{J}$ and $\tau=Jt$.
But if we change the initial condition, the solutions changes. For the case where $c_1(0) = 0$, $c_2(0) = 1$ and $c_3(0) = 0$, the solutions are:
\begin{eqnarray}
c_1(t) &=&  \frac{\sqrt{2}i e^{-\frac{\bar{U}\tau}{2}}}{\bar{\Omega}}\hspace{2mm} \sin(\bar{\Omega}\tau)\nonumber \\
c_2(t) &=& \frac{e^{-i\bar U\tau}}{2} + \frac{e^{-\frac{i\bar U\tau}{2}}}{2}[\cos(\bar{\Omega}\tau) - \frac{\bar{U'}}{2\bar{\Omega}}\sin(\bar{\Omega}\tau)] \nonumber\\
c_3(t) &=& -\frac{e^{-i\bar U\tau}}{2} + \frac{e^{-\frac{i\bar U\tau}{2}}}{2}[\cos(\bar{\Omega}\tau) - \frac{\bar{U'}}{2\bar{\Omega}}\sin(\bar{\Omega}\tau)] 
\end{eqnarray}

\subsection{Three fermions}\label{3.5}

The wave function is given by,
\begin{eqnarray}
 |\psi(t)\rangle = c_1 (t) |\uparrow \downarrow, \uparrow\rangle + c_2(t) |\uparrow \downarrow, \downarrow\rangle + c_3(t) |\uparrow, \uparrow \downarrow\rangle +c_4(t) |\downarrow, \uparrow \downarrow\rangle  \nonumber
\end{eqnarray}
The state $|\uparrow \downarrow, \uparrow\rangle$ represents two fermions (one with up spin and the other with down spin) are in the left well and one fermion (with up spin) is in the right well and $|c_1(t)|^2$ is the probability of finding the state and so on for the other states. Substituting the above equation in the time-dependent Schr\"{o}dinger equation one readily finds that the equation of motions of $c_1$ and $c_3$ form a pair of closed coupled equations while those of $c_2$ and $c_3$ constitute a separate pair of closed coupled equations. Thus we have 
\begin{eqnarray}
i\frac{dc_i(t)}{dt} = -Jc_j(t) + \left ( U_0 + 2U_{lr} \right ) c_i(t) 
\end{eqnarray}
where $i\ne j$ stands for any of paired indexes. The solutions are given by 
\begin{eqnarray} 
c_1(t)=c_2(t) = \frac{e^{-i(U+2U_{lr})t}}{\sqrt{2}} \cos(Jt)\nonumber\\
c_3(t)=c_4(t) = i\frac{e^{-i(U+2U_{lr})t}}{\sqrt{2}} \sin(Jt)
\end{eqnarray}
subject to the initial conditions $c_1(0) = \frac{1}{\sqrt{2}}$, $c_2(0) = \frac{1}{\sqrt{2}}$, $c_3(0) = 0$ and $c_4(0) = 0$. For this case, we have  $\Delta C^F = 1/2$, $\Delta S^F = 0$, $\Delta W^F = \sin(2 J t)$ and $ \Delta^F_{{\rm SQL}} = \sin(2Jt)/12$. Thus we obtain 
\begin{equation}
\Sigma_p = {(\Delta C^F)^2+(\Delta S^F)^2} - \Delta_{{\rm SQL}} = \frac{1}{4} \left [ 1  - \frac{1}{3} \sin(2Jt) \right ] 
\end{equation}
\begin{equation}
\Sigma_w = (\Delta W^F_n)^2 - \Delta_{{\rm SQL}} = \frac{1}{3} \left [ \frac{1}{3} \sin^2(2 Jt) - \frac{1}{4} \sin(2Jt) \right ]
\end{equation}
These equations clearly show that for all times $\Sigma_p > 0 $, implying that there is no phase squeezing in this 3-fermion case. The number squeezing occurs when $\sin(2 J t) < 3/4$. Furthermore, the interaction term appears in the overall or global phase of the wave function and so all physical observables are independent of the interaction and depend only on $J$.

\section{Results and discussions}
In a previous paper \cite{jpb:2013:biswajit}, it was shown that, for low number of bosons,  the unitary phase operators defined by equations (\ref{eq6},\ref{eq7}) yield results which are quite different from those given by the non-unitary Carruthers-Nieto type operators  while for matter-waves with large number of bosons (as in BEC)  the results for the two kinds of operators tend to converge. This fact establishes the importance of the vacuum fluctuations in a few-boson quantum systems such as the DW systems considered here. The effects of contact interaction on the quantum phase dynamics of a few bosons and a pair of  fermions in a symmetric DW potential have been studied previously \cite{jpb:2013:biswajit}. In this paper, we are primarily interested in the effects of finite range of two-body interactions on the quantum dynamics and phase properties of bosons and two or three fermions in a symmetric DW potential. For the symmetric DW  potential of the equation (\ref{eq18}), the effects of the range $r_0$ 
of the two-body interaction of equation (\ref{eq22}) for the $s$-wave scattering length $a_s \rightarrow \pm \infty$ on the on-site interaction $U_{i}$ (subscript $i$ stands for either site-index `1'(right) or `2'(left)) is shown in figure 1. Note that here we have scaled all the length quantities by the axial harmonic oscillator length scale $a_{l}$ in the harmonic or tight-binding approximation of the single well of the DW potential. For symmetric DW well, we have $U_{l} = U_{r}$ and let us then  denote the common on-site interaction by  $U_0$ ($ =U_{l} = U_{r}$). Accordingly, all the energies are scaled by the corresponding harmonic oscillator ground-state energy. In these units, the fixed parameters of the figure 1 are chosen as $\bar\lambda = 0.5, \bar\eta = 2, \omega_{\rho} = 10\omega_z$, thus the height of the barrier in DW potential is $\bar V_0 = 2$. Figure (\ref{Figure 1.}) shows that as a function of the range of the interaction, there is a maximum at $r_0 = 0.07a_l$ and the maximum value of the on-site interaction is about 0.2. The exact numerical solution of the DW potential by DVR method yield the energies of the two lowest 
quasi-degenerate levels as $0.900$ and $0.948$. The eigenenergy of the next higher level is $2.18$. These numbers clearly indicate that we are well within the tight-binding approximation. In the limit $r_0 \rightarrow 0$, $U_0$ reduces to a small value while in the limit $r_0 \rightarrow \infty$ it goes to zero. The single-particle tunneling matrix element $J$ is calculated to equal to 0.024 which is fixed for all our numerical results. For all our numerical results, the scattering length $a_s$ is large ($|a_s| >\!>1$),  and for the model interaction potential chosen, corresponds to the unitarity regime where results are insensitive to any change of the scattering length.

\begin{figure}[b]
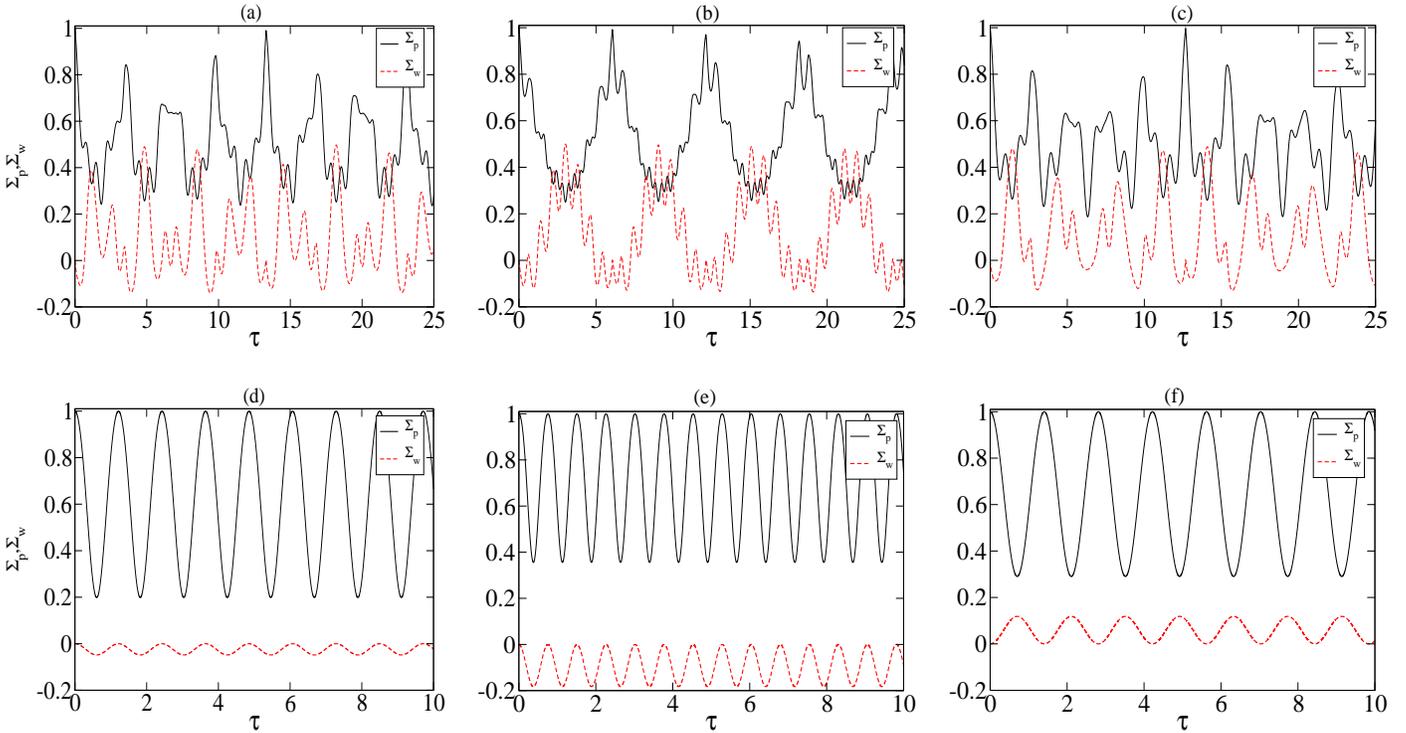

\centering
\vspace{.25in}
  \begin{tabular}{@{}cccc@{}}
    \hspace{-.4in}
    \includegraphics[height=1.8in, width=2.3in]{seq1.eps} &
    \includegraphics[height=1.8in, width=2.3in]{seq2.eps} &
    \includegraphics[height=1.8in, width=2.3in]{seq3.eps}\\
    \vspace{-0.2cm}\\  
    \hspace{-.4in}
    \includegraphics[height=1.8in, width=2.3in]{seq4.eps} &
    \includegraphics[height=1.8in, width=2.3in]{seq5.eps} &
    \includegraphics[height=1.8in, width=2.3in]{seq6.eps}  
  \end{tabular}
 \caption{\small Plotted are the number and phase squeezing parameters $\Sigma_w$ and $\Sigma_p$, respectively; as a function of dimensionless time $\tau = J t$ for two-boson system for $r_0 = 0.01 a_l$ (a,d), $r_0 = 0.1 a_l$ (b,e) and $r_0 = 0.5 a_{l} $ (c,f) with initially both the particles in right well (a,b,c) and with initially each particle in each well (d,e,f). Here the tunneling coefficient $\bar J=0.024$ (in unit of $\hbar \omega_z$). The parameter $r_0 = 0.01$, $r_0 = 0.1$ and $r_0 = 0.5$ correspond respectively to the on-site interaction  $U_0 = -0.08$, $U_0 = -0.17$ and $U_0= -0.04$ while the inter-site interaction $U_{12}$ is two orders of magnitude smaller than $U_0$ for all the three cases. Therefore, all the plots correspond to the strong interaction regime $U_0/J > 1$.}
  \label{Figure 2.}
\end{figure}

Here we present and analyze numerical results for the quantum dynamics and quantum fluctuation properties such as  number-phase uncertainty, number- and phase-squeezing of both bosons and fermions.  We mainly focus on the evolution of the fluctuation properties of the 
unitary quantum phase-difference and number-difference operators, and the effects of the interaction range at large scattering length on these properties. In what follows we first describe the results for few-boson systems and then two- and three-fermion systems.

\begin{figure}[b]
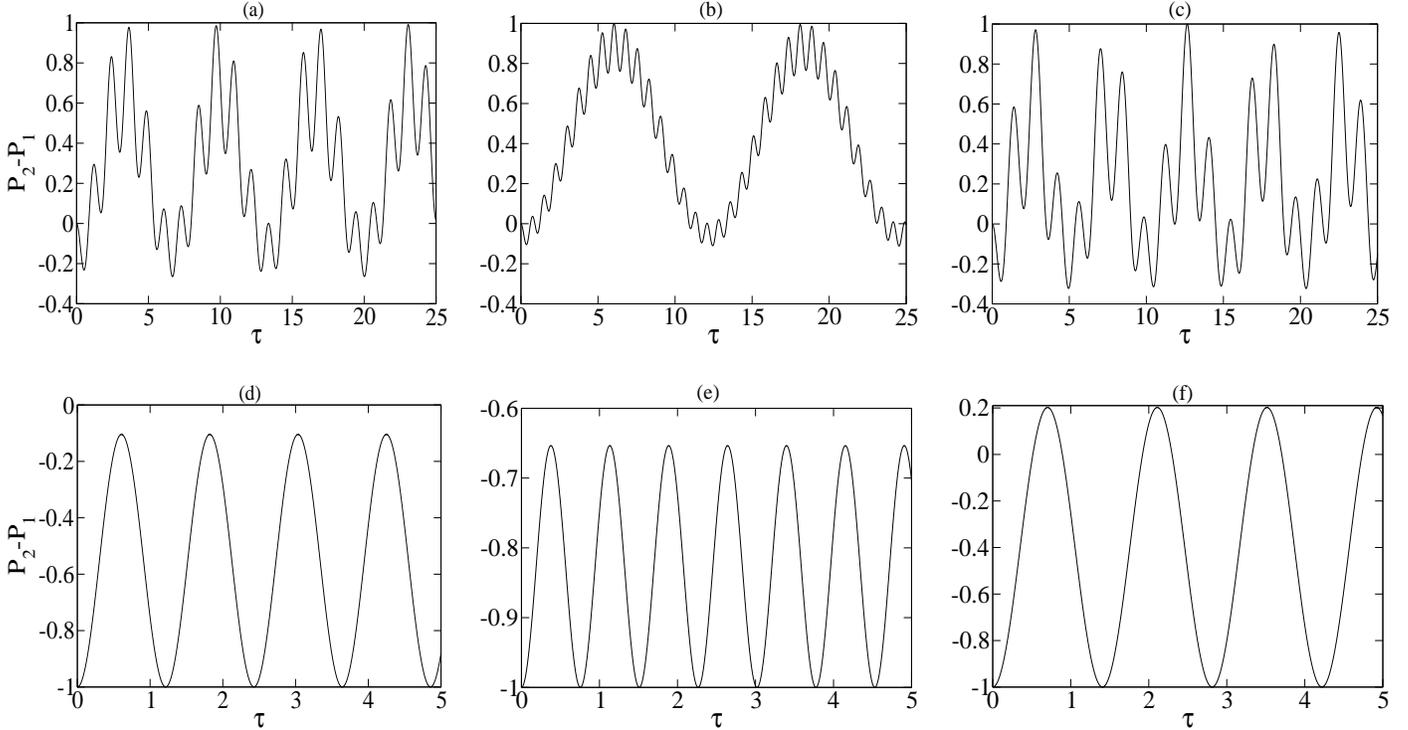

\centering
\vspace{.25in}
  \begin{tabular}{@{}cccc@{}}
    \hspace{-.4in}
    \includegraphics[height=1.8in, width=2.3in]{tun1.eps} &
    \includegraphics[height=1.8in, width=2.3in]{tun2.eps} &
    \includegraphics[height=1.8in, width=2.3in]{tun3.eps}\\
    \vspace{-0.2cm}\\  
    \hspace{-.4in}
    \includegraphics[height=1.8in, width=2.3in]{tun4.eps} &
    \includegraphics[height=1.8in, width=2.3in]{tun5.eps} &
    \includegraphics[height=1.8in, width=2.3in]{tun6.eps}  
  \end{tabular}
  \caption{\small Time evolution of the difference between probabilities of the both the particles being  in the left well ($P_2$)  and single-particle occupancy ($P_1$) for two-boson system for the same parameters as in figure (\ref{Figure 2.}), i.e.,the parameters of the subplots (a), (b), (c), (d), (e) and (f) are the same as those of  the subplots (a), (b), (c), (d), (e) and (f), respectively, of figure (\ref{Figure 2.}). }
  \label{Figure 3.}
\end{figure} 
  
\begin{figure}[b]
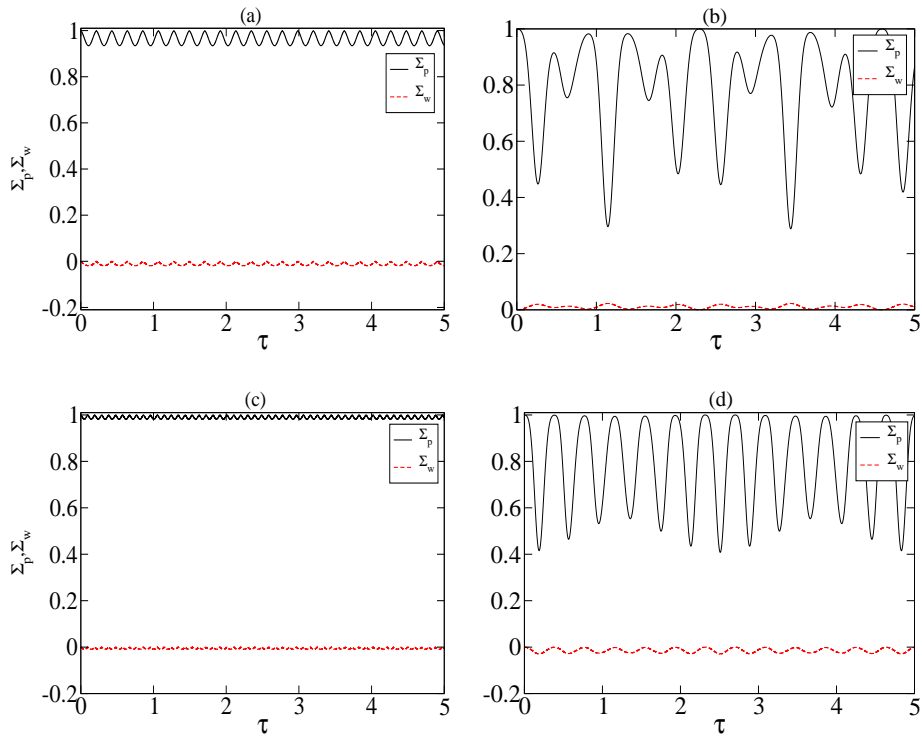

\centering
\vspace{.25in}
  \begin{tabular}{@{}cccc@{}}
    \hspace{-.4in}
    \includegraphics[height=1.8in, width=2.3in]{10sa.eps} &
    \includegraphics[height=1.8in, width=2.3in]{5v5.eps}\\
     \vspace{-0.2cm}\\  
    \hspace{-.4in}
     \includegraphics[height=1.8in, width=2.3in]{10sa2.eps} &
    \includegraphics[height=1.8in, width=2.3in]{5v52.eps}
     \end{tabular}
 \caption{\small Time evolution of squeezing parameters for $N=10$ boson system for $r_0 = 0.01 a_l$ (a,b) and $r_0 = 0.1 a_l$ (c,d) with initially all the particles in right well (a,c) and with initially equal number of particles initially in each well (b,d).}
 \label{Figure 4.}
\end{figure}

\subsection{Bosons} 
  
  Figure (\ref{Figure 2.}) shows the time evolution of the squeezing parameters $\Sigma_p$ and $\Sigma_w$ for $N=2$  bosons  for three different ranges $r_0=0.01$, $r_0=0.1$ and $r_0=0.5$.   From this figure, we notice that, when initially both bosons are in the same well (upper panel in figure 2), the temporal evolution of the normalized  number fluctuation $\Delta W_n$ exhibits periodic squeezing ( $\Sigma_w < 0$) with multi-periodicity  while phase fluctuation is always above the SQL ($\Sigma_p>0 $) for all three ranges. Phase fluctuation also shows multi-periodic behavior with the largest reduction in fluctuation being about 70\% of the maximum value of unity.   In contrast, when the initial condition is changed to each boson being in each well (lower panel of figure \ref{Figure 2.}), we notice that both the number and phase fluctuations oscillate periodically with almost a single period. The subplots 2(d) and 2(e) show that $\Delta W_n$ oscillate below or at most at the level of SQL ($\Sigma_w \le 0$) 
implying that the number 
fluctuation is squeezed at most of the time. However, in subplot 2(f) we notice that the $\Delta W_n$ oscillate above the SQL. This contrasting behavior for different ranges can be explained as due to  the on-site interaction. The on-site interaction in 2(f) is much smaller than that  in subplots 2(d) and 2(e) while it is much larger in subplot 2(e). Squeezing in number variables requires nonlinearity in the Hamiltonian in terms of number operator, here the nonlinearity is given by the interaction. In subplot 2(d), the $\Delta W_n$ is periodically and marginally squeezed while in subplot 2(e) it is largely squeezed. In accordance with the principle of uncertainty, we observe that the  phase fluctuation at the same time is maximum in subplot 2(d) among the three subplots of the lower panel in figure 2. 

Now, on the question of why two different initial conditions give rise to two almost completely different features in fluctuation properties has to do with the single- and  double-occupancy or pair-
tunneling  and 
memory effect of the systems. In figure (\ref{Figure 3.}) we have plotted the difference $P_2 - P_1$ between the tunneling probability ($P_2$)  and the single-occupancy ($P_1$) as a function of time for the same ranges and the initial condition as in figure (\ref{Figure 2.}). Here $P_2$ is defined as the probability of finding the two particles in the left well when both of them were initially in the right well.  Note that $P_2$ is different from the double occupancy or pair-probability $P_{pair}$ which is the total probability of finding both the particles in the same well \cite{selim:prl:2015}. We notice that  $P_2$  dominates over $P_1$ for most of the times (subplots 3(a),(b) and (c)). 
 By comparing the figures (\ref{Figure 2.}) and (\ref{Figure 3.}), one can notice that,  larger the $P_2$ is, the larger is the $\Delta W_n$ and smaller is the phase fluctuation at any instant of time. As a consequence, in the initial condition of each boson being in each well, $\Delta W_n$ is mostly squeezed provided the interaction is strong enough (figure 2(e)). We have found that, for a small interaction time $t$ such that $U_0 t/\hbar <\!< 1$,   $P_{pair}$  decreases while $P_1$ increases as a function of the on-site interaction.           
  
 Next, in figure (\ref{Figure 4.}) we show the time evolution of the two squeezing parameters with $N= 10$ bosons for two different initial conditions: (1) all 10 bosons are initially in the same well and (2) 50\% of the bosons, that is, 5 bosons are initially in each well. Here we have chosen two ranges $r_0 = 0.01$ and 
 $r_0 = 0.1$ which correspond to smaller and larger on-site interaction, respectively. For $r_0 = 0.01$,   
 we observe that, for the former initial condition the $\Delta W_n$ oscillates with very small amplitude around zero  while the phase fluctuation oscillates with small amplitude close to unity. In the latter initial condition for $r_0 = 0.01$, $\Delta W_n$ oscillates above zero implying there is no number squeezing. In this case phase fluctuation is largely reduced albeit above  SQL. Now, when the range is changed to $r_0=0.1$, that is, corresponding to larger on-site interaction, $\Delta W_n$ oscillate periodically just below the SQL in the latter initial condition. When we compare these results with those for $N=2$ bosons in figure (\ref{Figure 2.}), we notice that for $N = 10$ bosons,  $\Delta W_n$ is less squeezed. We have checked that in the limit of large bosons $\Delta W_n$ tends to settle down at the SQL coherent level as discussed in subsection (\ref{2.1}) while the phase fluctuation $\Delta E_{\phi}$ is close to  unity. We have also checked that for large number of bosons, the contribution of 
vacuum terms $C_{12}^{(0)}$ and $S_{12}^{(0)}$ tend to vanish while the quantity $\sqrt{(C_{12})^2 + (S_{12})^2} \rightarrow 1$. This means that $\Delta W \rightarrow \sqrt{N}$ in the limit $N \rightarrow \infty$.

  \subsection{Fermions}
\begin{figure}[h]
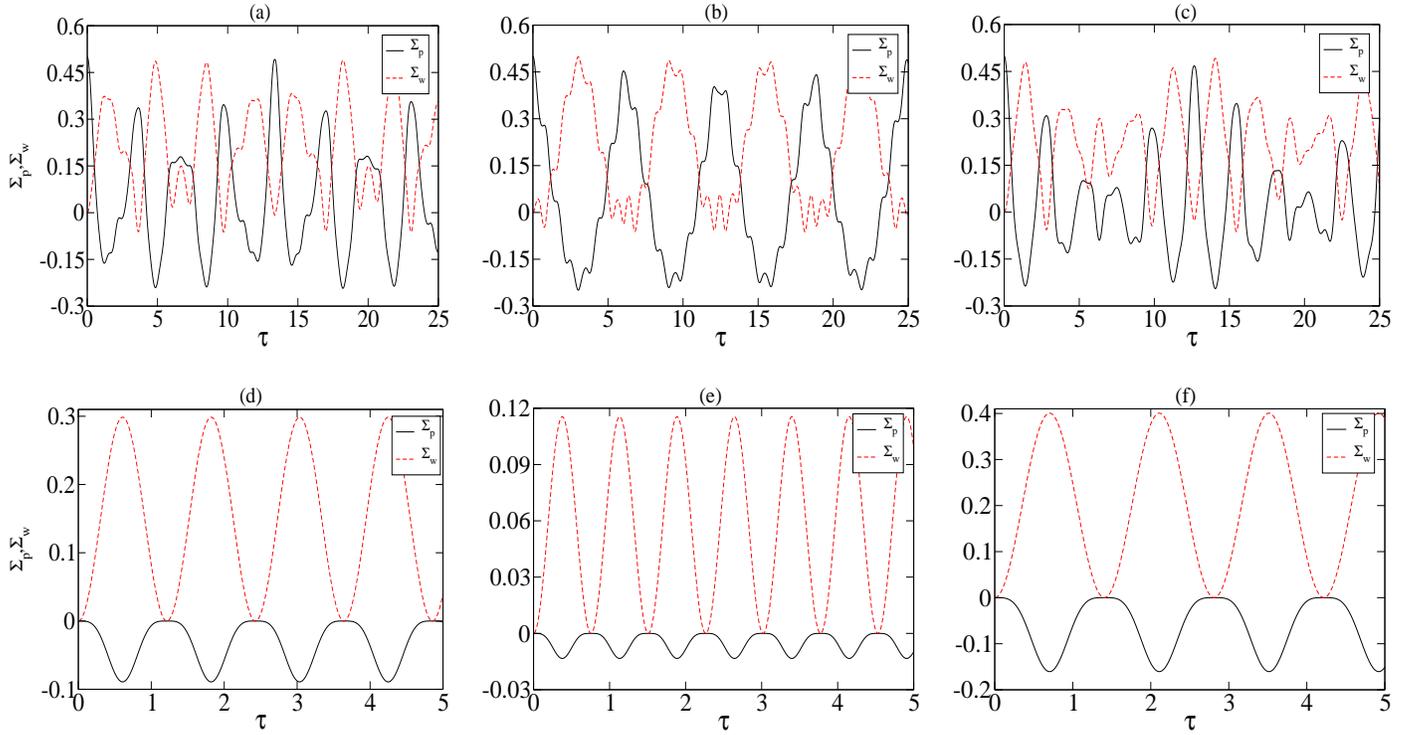

\centering
\vspace{.25in}
  \begin{tabular}{@{}cccc@{}}
    \hspace{-.4in}
    \includegraphics[height=1.8in, width=2.3in]{fer1.eps} &
    \includegraphics[height=1.8in, width=2.3in]{fer2.eps} &
    \includegraphics[height=1.8in, width=2.3in]{fer3.eps}\\
    \vspace{-0.2cm}\\  
    \hspace{-.4in}
    \includegraphics[height=1.8in, width=2.3in]{fer4.eps} &
    \includegraphics[height=1.8in, width=2.3in]{fer5.eps} &
    \includegraphics[height=1.8in, width=2.3in]{fer6.eps}  
  \end{tabular}
  \caption{\small Time evolution of squeezing parameters for two-fermion system for $r_0 = 0.01 a_l$ (a,d), $r_0 = 0.1 a_l$ (b,e) and $r_0 = 0.5 a_{l} $ (c,f) with initially both the particles in right well (a,b,c) and with initially each particle in each well (d,e,f).}
  \label{Figure 5.}
\end{figure}

  Here we calculate phase and number fluctuations of spin-half fermions. We consider phase difference between the two localized fermionic states, namely either spin up ($\uparrow$) or down ($\downarrow$) state localized in one well and spin  down ($\downarrow$) or up ($\uparrow$) state,  respectively, in the other well.   
 In figure (\ref{Figure 5.}), we display the variation of $\Sigma_w$ and $\Sigma_p$  as a function of time for  $N=2$ spin-half fermions for three different ranges and two initial conditions as in figure (\ref{Figure 2.}). We here observe completely opposite trend of fluctuation behavior as compared to  that in two-boson system. Phase fluctuation shows  significant squeezing while  number fluctuation lies mostly above the SQL. In contrast, we earlier noticed that there is no phase squeezing in bosonic case. However, in case of three fermions, phase fluctuation does not exhibit squeezing but the number fluctuation can go below SQL.  
 
 The  phase squeezing for $N=2$ two-component fermions may be due to fermionic symmetry. Since we are considering $s$-wave interaction, the spin-state of the two fermions is singlet. Therefore, the spatial part of the wavefunction of two fermions must be symmetric. So, the state with one spin  $\uparrow$ fermion in the right well and the other spin $\downarrow$ fermion in the left well and the state with one spin  $\downarrow$ fermion in the right well and the other spin $\uparrow$ fermion in the left well form a two-particle superposition state of symmetric combination and therefore the two states are entangled. This leads to phase squeezing which can also be regarded as a manifestation of spatial exchange  symmetry or spatial entanglement. In fact, it has been experimental demonstrated that the atomic interferometry beyond classical limit in a Bose-Einstein condensate requires spin-dynamics and pair-entanglement \cite{science:2011}. It is worth-mentioning that average values of number 
and phase operators and other average quantities such as single- and double-occupancy for two bosons and two two-component fermions are similar. So, by measuring these average quantities one can not probably distinguish between the two-boson and two-fermion spin-half systems. It is the quantum fluctuation properties that have clearly distinctive features. In the limit of on-site interaction going to zero, the periodicity of the phase or number fluctuation is $\pi/2$ as is evident from the subplots (c) and (f) of figure (\ref{Figure 5.}). This result also follows from the analytical results discussed in subsection (\ref{3.4}). However, when the on-site interaction is finite, the periodicity is less than $\pi/2$ as can be seen from the subplots (b) and (e).

 \section{Conclusions and outlook}
 
 In conclusion, we have studied number-phase uncertainty, number- and phase-squeezing of interacting bosons and fermions in a double well potential under two-mode approximation. The total number of particles $N$ in our system is a conserved quantity. By employing two phase-difference operators $\hat{C}_{12}$ and $\hat{S}_{12}$ corresponding to the measurement of the cosine and sine of the phase-difference between the modes `1' and `2' of bosonic or fermionic matter-waves, we have established an uncertainty relation for the product of the fluctuation of the normalized number-difference $\Delta W_n$ and a mean phase fluctuation $\Delta E_{\phi}= \sqrt{ (\Delta C_{12})^2 + (\Delta S_{12})^2}$, where $\Delta C_{12}$ and $\Delta S_{12}$ are the fluctuations in $ \hat{C}_{12}$ and $\hat{S}_{12}$, respectively. Accordingly, we have defined a standard quantum limit (SQL) or shot noise for both  $\Delta W_n$ and $\Delta E_{\phi}$. Both phase operators and the SQL depend on the coupling between the vacuum state (empty 
mode) of one mode and the maximally occupied other mode. This vacuum coupling is introduced to fulfill the unitarity condition of the phase operators. For small number of bosons, this vacuum coupling has significant effect on the fluctuation properties while  for large number of bosons the effect of vacuum coupling diminishes.     
 
 To study the effect of two-body interaction on quantum fluctuation and dynamics, we have used a finite-range model interaction potential which depends on two parameters, namely the range $r_0$ and the $s$-wave scattering length $a_s$.  Our model potential is valid for $|a_s| > 2 r_0$ and so is more suitable for large scattering length and so can take into account the effects of scattering resonances such as Feshbach resonances.  The finite-range interaction can lead to not only on-site interaction, but also small inter-site interaction. Since for neutral non-polar cold atoms, the interaction is usually of extremely small range, we do not consider large range or long range interaction. For such finite-range interaction, it is basically on-site interaction which dominates in the two-mode or tight-binding approximation. We have demonstrated the effects of $r_0$ on the quantum fluctuations of the two-mode number- and phase-difference operators when $|a_s| >\!> r_0$. In this limit of large scattering length, the 
results show universal behavior in the sense that they do not depend on $a_s$ but depend only on $r_0$. Our results show that, depending on the initial condition, the range of interaction has significant effect on the number-squeezing, phase fluctuation  and quantum dynamics of bosons. In particular, for the ranges at which the on-site interaction is large, we have found significant reduction or squeezing of number fluctuation at times when the bosons are more or less evenly distributed into the two sites of the double well.
 
 Unlike that in bosons, the phase fluctuation of two fermions exhibit squeezing. This may follow from Pauli's exclusion principle shielding the fermions from occupying states that are already occupied. Recently, phase fluctuation below the shot-noise in a two-component Bose-Einstein condensate (BEC) has been experimentally demonstrated by ``superfluid shielding'' of one of the components \cite{prl:2016:ketterle}. The underlying physical mechanism of phase squeezing in fermions and two-component BEC with ``superfluid shielding'' may be related to the anti-correlation.  Furthermore, 
 the fact that the sub-shot noise phase fluctuation is experimentally found to be robust \cite{prl:2016:ketterle} when the number of bosons in either component is small suggests a unitary quantum phase operator-based method is necessary to calculate the  phase and its fluctuation of small or   mesoscopic quantum systems.  
 
 Our formalism for number-phase uncertainty for bosonic and fermions matter-waves, and the effects described herein of the resonant interactions on the quantum fluctuations of number and phase operators may find applications in hitherto-unexplored matter-wave interferometry with few bosonic or fermionic atoms. Particularly interesting will be the the study of quantum dynamics for coupled phase and number operators in Josephson oscillations in mesoscopic systems such as finite number of interacting bosons or fermions in a double well. Although a pair of interacting bosons or two-component fermions trapped in a DW potential can be  regarded as a building block for Bose-Hubbard or Fermi-Hubbard model, respectively, the average values of physical quantities  will show similar qualitative behavior for both the bosonic and fermionc cases, but the quantum fluctuations in two cases will be quite different. So, it would be an interesting experimental pursuit to measure the fluctuations of  quantum phase operators 
for two particles trapped in a DW potential in both bosonic and fermionic cases. 

Reduced quantum phase fluctuation is a key to high-precision interferometric measurements. For instance, Laser Interferometer Gravitational-Wave Observatory (LIGO) \cite{ligo}  makes use of laser's reduced phase fluctuation at the shot-noise limit.
A couple of years back, LIGO  made the first successful observation of gravitational waves,  for which this year's   Nobel prize in physics has been awarded to Rainer Weiss,  Barry C. Barish and Kip S. Thorne.  Future research should explore  methods to achieve two-mode 
quantum phase-squeezed optical fields which  may be useful for making sub-shot noise optical interferometers to detect gravitational waves with higher precision.
  
\vspace{0.5cm}

\noindent
{\bf Acknowledgment}\\
 This work is supported by the project SB/S2/LOP-008/2014 of the Department of Science \& Technology, Govt. of India. One of us (SM) is thankful to the Council of Scientific and Industrial Research, Govt. of India, for a support.

\end{document}